\documentclass[PoP,twocolumn,showpacs,%
preprintnumbers,amssymb,prb]{revtex4-1}
\usepackage{graphicx}
\usepackage[colorlinks=true,linkcolor=blue,%
linkcolor=blue,pagecolor=blue,citecolor=blue,%
urlcolor=blue,anchorcolor=black,%
bookmarksnumbered=true,%
bookmarksopen=true]{hyperref}
\usepackage{dcolumn}
\usepackage{bm}
\renewcommand{\frac}[2]{\displaystyle{#1 \over #2}}
\begin{document}
\title{Measurement of the speed of sound by 
observation of the Mach cones in a complex 
plasma under microgravity conditions}
\author{D.~I.~Zhukhovitskii} \email{dmr@ihed.ras.ru}
\author{V.~E.~Fortov}
\author{V.~I.~Molotkov}
\author{A.~M.~Lipaev}
\author{V.~N.~Naumkin}
\affiliation{Joint Institute of High Temperatures, Russian 
Academy of Sciences, Izhorskaya 13, Bd.~2, 125412 
Moscow, Russia}
\author{H.~M.~Thomas}
\affiliation{Research Group Complex Plasma, DLR, 
Oberpfaffenhofen, 82234 Wessling, Germany}
\author{A.~V.~Ivlev}
\affiliation{Max-Planck-Institut f\"{u}r extraterrestrische 
Physik, Giessenbachstrasse, 85748 Garching, Germany}
\author{M.~Schwabe}
\affiliation{Department of Chemical and Biomolecular 
Engineering, Graves Lab, D75 Tan Hall, University of 
California, Berkeley, CA 94720, USA}
\author{G.~E.~Morfill}
\affiliation{Max-Planck-Institut f\"{u}r extraterrestrische 
Physik, Giessenbachstrasse, 85748 Garching, Germany}

\begin{abstract}
We report the first observation of the Mach cones excited by a larger 
microparticle (projectile) moving through a cloud of smaller 
microparticles (dust) in a complex plasma with neon as a buffer gas 
under microgravity conditions. A collective motion of the dust 
particles occurs as propagation of the contact discontinuity. The 
corresponding speed of sound was measured by a special method of 
the Mach cone visualization. The measurement results are 
incompatible with the theory of ion acoustic waves. The estimate for 
the pressure in a strongly coupled Coulomb system and a scaling law 
for the complex plasma make it possible to derive an evaluation for 
the speed of sound, which is in a reasonable agreement with the 
experiments in complex plasmas.
\end{abstract}
\maketitle

\section{\label{s1} Introduction}

Dusty or complex plasma is a low-temperature plasma, 
which includes dust particles with sizes ranging from $1$
 to $10^3 \;\mu {\mbox{m}}$. Due to the higher electron 
mobility, particles acquire a considerable electric charge. 
Thus, a strongly coupled Coulomb system is 
formed.\cite{1,2,3,4,5,6,8,9} In such plasma, various 
collective phenomena at the level of individual particles 
take place. Complex plasmas are usually studied in gas 
discharges at low pressures, e.g., in the radio frequency 
(RF) discharges. Under microgravity conditions, a large 
homogeneous bulk of the complex plasma can be observed. 
The particles form a nearly homogeneous cloud around the 
center of the chamber, typically with a central void caused 
by the ions streaming outwards. The microgravity 
conditions are realized either in parabolic 
flights\cite{10,11,12,13,14} or onboard the International 
Space Station (ISS).\cite{10,15,16,17,18,019,19}

Some experiments are carried out on an inhomogeneous 
system consisting of the particles with different diameters. 
The simplest example of such a system is a large particle 
surrounded by a dense cloud of smaller particles. Usually, 
this particle called the projectile moves through the cloud 
with a supersonic or subsonic velocity. Such projectiles are 
generated using controlled mechanisms of 
acceleration,\cite{11,20} or they can appear 
sporadically.\cite{19,21} In the latter case, agglomerates or 
larger particles left over from previous experiments (not 
removed during the cleaning procedure and accumulated at 
the periphery of the particle cloud) are cracked upon 
illumination by the laser sheet or upon shaking the 
chamber. Detached individual particles acquire a negative 
charge and then they are accelerated due to the Coulomb 
repulsion.\cite{22}

In the work by Havnes {\it et al.\/},\cite{25} propagation of 
a long-wave nondispersive disturbance, which is usually 
called the sound, and formation of a cone corresponding to 
the Mach cone in a continuous medium were predicted for 
the strongly coupled system of the dust particles. In the first 
experiments, the Mach cones in the 2D plasma crystal were 
excited by a sphere moving faster than the lattice sound 
speed beneath the 2D lattice plane (Samsonov {\it et 
al.\/}\cite{26,21}) and by applying a force from the 
radiation pressure of a moving laser beam (Melzer {\it et 
al.\/}\cite{30}). Later, Nosenko {\it et al.\/}\cite{27,28} 
used the latter method of the disturbance excitation and 
observed several Mach cones, which were attributed by the 
authors to propagation of the compressional and shear 
wakes and to their interference. The shape of the Mach 
cones formed by nondispersive linear sound waves was 
calculated analytically, and the curved wings of the Mach 
cones were experimentally observed by Zhdanov {\it et 
al\/}.\cite{31,32}

In the recent works by Caliebe, Arp, and Piel,\cite{11} 
Jiang {\it et al.\/}\cite{019}, and Schwabe {\it et 
al.\/},\cite{19} the excitation of the 3D Mach cones by the 
supersonic projectiles moving in a strongly coupled cloud 
of charged particles was observed. In these studies, argon 
was used as a buffer gas. The determined speed of sound is 
not much different in the performed experiments.

In this work, we report the first measurement of the speed 
of sound in the dust cloud by observation of the Mach 
cones in the case that neon is used as a buffer gas. The 
measured speed of sound proved to be more than twice as 
low as for argon, which means that this quantity depends 
rather sensitively on the sort of a gas. We can account for 
observed dependences on the basis of a scaling law for the 
dust cloud obtained in Ref.~\onlinecite{22}.

The paper is organized as follows. In Sec.~\ref{s2}, we 
describe the experimental setup and the method of the 
Mach cone visualization. The details of experimental data 
processing and the results of speed of sound measurement 
are presented in Sec.~\ref{s3}. A qualitative interpretation 
for our experiment is discussed in Sec.~\ref{s4}, and the 
results of this study are summarized in Sec.~\ref{s5}.

\section{\label{s2} Experiment}

The experiment was performed during the 13th mission of 
PK-3 Plus on the ISS. The setup is described in detail in 
Ref.~\onlinecite{18}. The heart of this laboratory consists 
of a capacitively coupled plasma chamber with circular 
electrodes of 6~cm diameter and 3~cm apart. A radio 
frequency (RF) voltage applied to these electrodes 
generates a bulk of plasma. A dust cloud was formed by the 
microparticles injected into the main plasma with 
dispensers. Neon was used as a buffer gas at the pressures 
of 15 and 20~Pa, and the main microparticle cloud was 
composed of the monodisperse silica particles with the 
diameter of $1.55\;\mu {\mbox{m}}$. The diameter of 
observed projectiles estimated as the diameter of larger 
particles, which are also present in the chamber, is most 
likely the same as in the experiment\cite{19}, i.e., it is 
equal to $15\;\mu {\mbox{m}}$. The trajectories of the 
dust particles and the projectile were monitored using the 
optical particle detection system, which consisted of a laser 
illumination system and a recording system, containing 
three progressive scan CCD-cameras. The illumination 
system is based on two laser diodes with $\lambda = 
686\;{\mbox{mm}}$
 and a continuous wave optical power of 40~mW, the light 
of which is focused to a thin sheet. This laser light sheet has 
a full width at half maximum of about $80\;\mu 
{\mbox{m}}$
 at the focal axis. The cameras with different magnifications 
and fields of view recorded the light scattered by the 
microparticles at $90^\circ $. To analyze the microparticle 
motion, we used three cameras with different fields of view 
and resolutions, which showed the entire microparticle 
cloud between the electrodes. The plasma glow was filtered 
out. The cameras follow the PAL standard with a resolution 
of $768 \times 576$~pixels. Each camera provides two 
composite time interlaced video channels with 25~Hz 
frame rate. Both video channels from one camera were 
selected for recording, so they were combined to a 50~Hz 
progressive scan video.
\begin{figure}
\includegraphics[width=9.44cm]{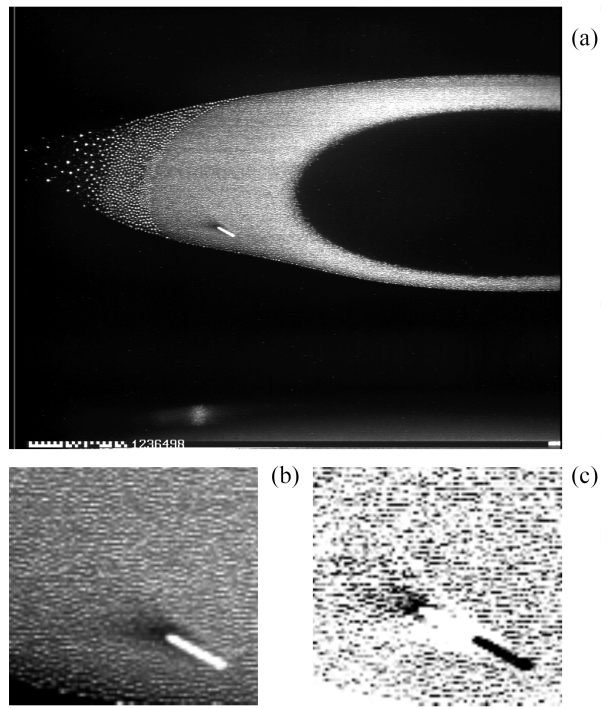}
\caption{\label{f1}(a) Snapshot of the projectile moving 
through the cloud of dust particles with a supersonic speed; 
(b) enlarged fragment of the snapshot; and (c) the result of 
the Mach cone visualization. The neon gas pressure is 
20~Pa.}
\end{figure}

We observed two events of the projectile motion through 
the dust cloud. Figure~\ref{f1} shows the first one. The 
images were recorded by the quadrant view camera with the 
resolution of $49.6\;\mu {\mbox{m}}$
 and $45.05\;\mu {\mbox{m}}$
 per pixel in horizontal and vertical directions, respectively, 
at 50~frames/s. The projectile moved with a supersonic 
velocity from the upper left to the lower right side of the 
dust cloud [Fig.~\ref{f1}(a)]. The track of the moving 
projectile is surrounded by a dust-free region (cavity), 
which emerges as a result of a strong Coulomb repulsion 
between the negatively charged dust particles and the 
projectile [Fig.~\ref{f1}(b)]. The cavity is elongated, the 
position of a projectile being eccentric. A comparison with 
[Fig.~\ref{f1}(c)] shows that in the center of a perturbation 
propagating through the dust cloud, the number density of 
dust particles is continuous both in the vicinity of the cavity 
and far apart from it. The perturbation proper has a typical 
form of the Mach cone. On this basis, we can conclude that 
observed perturbation is a \emph{contact 
discontinuity}.\cite{23}

Visualization of the Mach cone included the comparison of 
corresponding pixels for each pair of two successive video 
frames converted to 8-bit grayscale mode negative images. 
If a gray value for the latter image was within 10\% of that 
for the former image, a corresponding pixel of the resulting 
image was left blank [i.e., it is white in Fig.~\ref{f1}(c)]. 
Otherwise, the pixel assumed the value of the former 
image.

Although the dust particles form a strongly coupled system, 
they move around their equilibrium positions in the dust 
crystal. Typically, the positions of a dust particle in the 
successive frames differ by at least one pixel. Due to a low 
velocity, a point rather than a track in the image represents 
a particle. Consequently, many more of the unperturbed 
dust particles appear in the resulting image. However, if a 
particle finds itself in the contact discontinuity region, it is 
represented by a track due to a considerable velocity inside 
the perturbation. The tracks of neighboring particles 
overlap. Thus, a small region of the pixels with almost 
equal gray values is formed. Since the time interval 
between successive frames is 0.02~s, this region is shifted 
by 0.02~cm at the propagation velocity of 
$1\;{\mbox{cm/s}}$. If the perturbation thickness is larger 
than 0.02~cm (in our experiment, it is about 0.04~cm), its 
images in the successive frames partially overlap. This area 
of overlap is represented by the white pixels in the resulting 
image, which makes it possible to visualize the Mach cone 
in Fig.~\ref{f1} (obviously, the dust-free regions are also 
represented by the white pixels). Note that the efficiency of 
other methods used in Refs.~\onlinecite{11,019,19} would 
be insufficient for the visualization if neon was used as a 
carrier gas.

\section{\label{s3} Determination of the speed of sound}

It is well-known that the Mach angle $\theta$ is related to 
the speed of sound $c_s$ by the Mach cone relation $\sin 
\theta = c_s /u$, where $u$
 is the velocity of the perturbation source (in our case, this 
is the projectile velocity), i.e.,
\begin{equation}
c_s = u\sin \theta . \label{e1}
\end{equation}
The projectile velocity was determined by manual 
measurement of the positions of the projectile track centers 
in different frames. The velocity proved to increase from 
$4.6$
 to $5.8\;{\mbox{cm/s}}$
 as the projectile crossed the dust cloud. The estimates show 
that such acceleration along with a slight curvature of the 
projectile trajectory, which could lead to a bend of the 
rulings of a cone, would have a negligibly small effect on 
the result of determination of $c_s$ as compared to the 
measurement errors.

The Mach angle can be determined using the vectors 
${\bf{r}}_d$ and ${\bf{r}}_u$ that coincide with the lower 
and upper rulings of the Mach cone, respectively:
\begin{equation}
\sin \theta = \frac{1}{{\sqrt 2 }}\left( {1 - 
\frac{{{\bf{r}}_u \cdot {\bf{r}}_d }}{{r_u r_d }}} 
\right)^{1/2} . \label{e2}
\end{equation}
Alternatively, one can measure the Mach angle $\theta _d$ 
($\theta _u $) between the vector ${\bf{r}}_d$ 
(${\bf{r}}_u $) and the projectile displacement vector 
${\bf{s}}$, which coincides with the projectile track. Then
\begin{equation}
\begin{array}{*{20}c}
  {\sin \theta = \frac{{\sin \theta _u + \sin \theta _d }}{2},} 
\\
  {\sin \theta _{u,d} = \left[ {1 - \left( 
{\frac{{{\bf{r}}_{u,d} \cdot {\bf{s}}}}{{r_{u,d} s}}} 
\right)^2 } \right]^{1/2} .} \\
\end{array} \label{e3}
\end{equation}
\begin{figure}
\includegraphics[width=9.2cm]{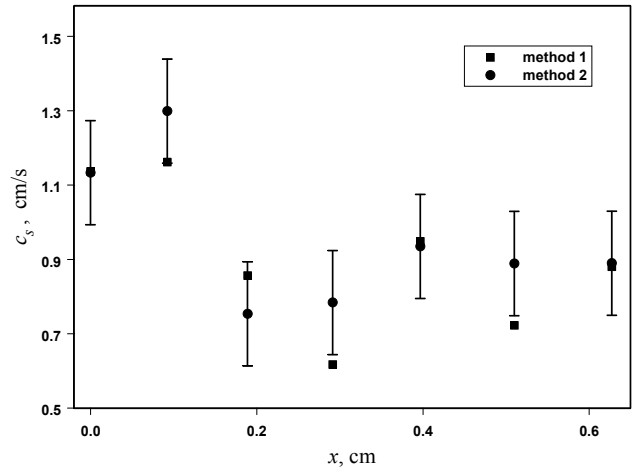}
\caption{\label{f2}Speed of sound as a function of the 
length of the projectile path in the dust cloud $x$
 (the first event). The methods of measurement are 
discussed in the text.}
\end{figure}

We determined the coordinates of the vectors ${\bf{r}}_d 
$, ${\bf{r}}_u $, and ${\bf{s}}$
 manually in each frame, which allowed one to measure 
$c_s $. Figure~\ref{f2} illustrates the results. 
\emph{Method 1} denotes the calculation by the formula 
(\ref{e2}); \emph{method 2} implies averaging of $\sin 
\theta$ given by Eqs.~(\ref{e2}) and (\ref{e3}). It is seen 
that both methods yield close results and no apparent 
dependence on the coordinate can be revealed within 
experimental errors. The averaging over the entire path of 
the projectile leads to the estimate $c_s = 0.96 \pm 
0.14\;{\mbox{cm/s}}$.

The second event of the projectile motion through the dust 
cloud was detected at the neon pressure of 15~Pa. The other 
parameters were the same as for the first event. Here, the 
Mach cone can be resolved only in three processed images, 
which increases the error. The average projectile velocity 
for this event $u = 2.4\;{\mbox{cm/s}}$
 is more than twice as low as for the above-discussed event, 
and the speed of sound still amounts to $c_s = 0.97 \pm 
0.51\;{\mbox{cm/s}}$, which is very close to the previous 
estimate. Thus, $c_s$ is more than twice as low as for 
argon\cite{11,019,19} revealing the effect of the gas sort.
\begin{figure}
\includegraphics[width=9.4cm]{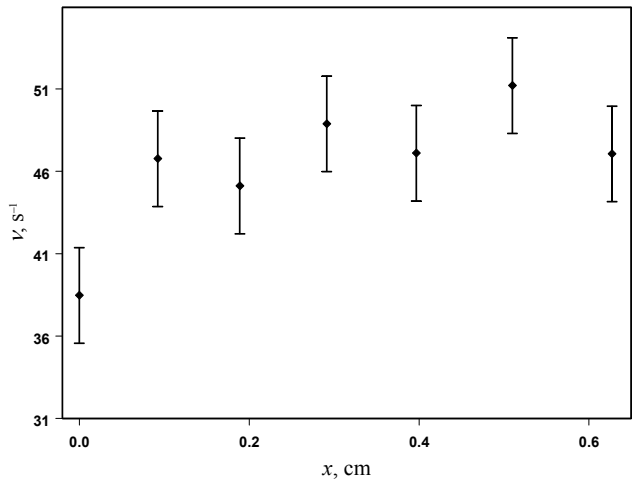}
\caption{\label{f3}Damping rate as a function of the length 
of the projectile path in the dust cloud $x$
 (the first event).}
\end{figure}

A clear resolution of the Mach cone rulings makes it 
possible to estimate the damping rate of the propagating 
perturbation as $\nu = (u/r)\cos \theta $, where $r$
 is the base radius of the Mach cone. Like $c_s $, $\nu$ 
reveals no apparent dependence on the coordinate along the 
projectile path (Fig.~\ref{f3}). Its average amounts to $\nu 
= 46 \pm 3\;{\mbox{s}}^{ - 1}$ for the first event and to 
$\nu = 32 \pm 13\;{\mbox{s}}^{ - 1} $, for the second one. 
For the first event, the damping length $l = c_s /\nu \approx 
0.021\;{\mbox{cm}}$, which is of the same order of 
magnitude as the visible perturbation wavelength $\lambda 
\approx 0.026\;{\mbox{cm}}$. Both lengths have the scale 
of ca.\ three to four interparticle distances.

\section{\label{s4} Discussion}

Under the conditions of our experiment, the volume charge 
of electrons is negligibly small as compared to that of the 
ions and particles,\cite{22} and the complex plasma can be 
treated as a system of negatively charged particles on the 
uniform positive background of the ions. Such situation is 
characteristic of strongly coupled Coulomb systems under 
high energy density. Similarly to the theory of ideal 
collisionless plasma, the perturbation treated in this work is 
commonly associated with the ion acoustic wave with the 
speed of sound\cite{24}
\begin{equation}
c_s = \left( {\frac{{\left| {Z_d } \right|T_d }}{{M_d }}} 
\right)^{1/2} , \label{e4}
\end{equation}
where $Z_d = a_d T_e \Phi _d /e^2$ is the dust particle 
charge in units of the elementary charge $e$, $a_d$ is the 
particle radius, $T_e$ is the electron temperature (the 
Boltzmann constant is set to unity), $\Phi _d = e\varphi 
/T_e $, $\varphi$ is the electrostatic potential of a particle, 
$T_d$ is the particle temperature that is related to the 
average kinetic energy of a particle, $M_d = (4\pi /3)\rho 
a_d^3$ is the particle mass, and $\rho$ is the density of the 
particle material. For our experiment, $T_e \simeq 
7\;{\mbox{eV}}$. If we use the orbital motion limited 
approximation\cite{9} for determination of $\Phi _d$ and 
set $T_d = T_n $, where $T_n = 300\;{\mbox{K}}$
 is the temperature of a buffer gas (room temperature), we 
arrive at the estimate $c_s = 10.2\;{\mbox{cm/s}}$. 
Obviously, this is yet a \emph{lower bound}. Thus, we can 
ascertain at least one order of magnitude disagreement 
between experiment and theory, which cannot be removed 
by existing alternative approaches to the calculation of the 
particle charge. Therefore, the formula (\ref{e4}) is 
\emph{fully inapplicable} for strongly coupled Coulomb 
systems.

For such systems, we will derive an alternative estimate for 
the speed of sound $c_s $. Obviously, $c_s^2$ is 
proportional to the ratio of the pressure, which is in the 
order of magnitude $Z_d^2 e^2 n_d^{4/3} $, to the mass 
density of a dust cloud $(4\pi /3)a_d^3 \rho n_d $. Hence,
\begin{equation}
c_s^2 \sim \frac{3}{{4\pi }}\frac{{e^2 n_i^2 }}{{\rho 
a_d^3 n_d^{5/3} }}, \label{e5}
\end{equation}
where $n_i$ is the ion number density and we used the 
quasineutrality condition $\left| {Z_d } \right|n_d \simeq 
n_i $. It was shown in Ref.~\onlinecite{22} that the overlap 
of potentials of the dust particles, which scatter streaming 
ions, leads to the scaling law for the dust cloud that relates 
the particle number density to the particle radius: $n_d^{ - 
2/3} = (4\pi /3)^{2/3} \kappa T_e a_d $, where $\kappa$ is 
some constant (the ``dust invariant"). We substitute this in 
(\ref{e5}) to derive
\begin{equation}
c_s \simeq \left( {\frac{{4\pi }}{3}} \right)^{1/3} (\kappa 
T_e )^{5/4} \frac{{en_i }}{{\rho ^{1/2} a_d^{1/4} }}. 
\label{e6}
\end{equation}

Unfortunately, neither $n_d$ nor $n_i$ are available for the 
experiment with neon. For this reason, we test the relation 
(\ref{e6}) on the experiments with argon (Table~\ref{t1}). 
It is seen that for $n_i = 5.5 \times 10^8 \;{\mbox{cm}}^{ - 
3} $, which is close to typical values for argon,\cite{22} 
this relation yields a reasonable agreement with the 
experiment and reproduces the observed dependences of the 
speed of sound on the particle radius and electron 
temperature. Equation (\ref{e6}) reproduces the determined 
speed of sound for neon ($0.96\;{\mbox{cm/s}}$) at $n_i = 
1.0 \times 10^8 \;{\mbox{cm}}^{ - 3} $, which also seems 
to be a reasonable value because the ion number density in 
neon is typically one order of magnitude lower than that in 
argon\cite{18} (however, $\kappa$ may be different for 
neon). Note that (\ref{e4}) disagrees with the experiments 
with argon as well, albeit the disagreement is not as great as 
for neon. For example, it leads to $c_s = 
4.4\;{\mbox{cm/s}}$
 for the experiment\cite{19}.
\begin{table}
\caption{Speed of sound $c_s$ in experiments with argon 
as a buffer gas at different argon pressures $p$
 and dust particle diameters vs.\ theoretical estimation 
Eq.~(\ref{e6}) at $\kappa = 0.209\;{\mbox{cm/eV}}$
\cite{22} and $n_i = 5.5 \times 10^8 \;{\mbox{cm}}^{ - 3} 
$.}
\label{t1}
\begin{ruledtabular}
\begin{tabular}{cccc}
$p$, Pa & $2a_d ,\;\mu {\mbox{m}}$
 & $c_s$ (exp.), ${\mbox{cm/s}}$
 & $c_s$ (\ref{e6}), ${\mbox{cm/s}}$
\\
\hline
9.6 & 1.55 & 2.4 (Ref.~\onlinecite{019}) & 2.26\\
10 & 2.55 & 2.2 (Ref.~\onlinecite{19}) & 2.18\\
30 & 9.55 & 2.0 (Ref.~\onlinecite{11}) & 2.14
\end{tabular}
\end{ruledtabular}
\end{table}

Consider the damping of a propagating perturbation. The 
damping due to the friction between the dust particles and a 
buffer gas (neutral or the Epstein drag\cite{33}), which 
always takes place in the complex plasma, is characterized 
by the friction coefficient\cite{6} $\nu _n = (8\sqrt {2\pi } 
/3)\delta m_n n_n v_{T_n } a_d^2 /M_d $, where $\delta 
\simeq 1.4$
 is the accommodation coefficient, $m_n$ is the mass of a 
buffer gas molecule, $n_n$ and $v_{T_n } = (T_n /m_n 
)^{1/2}$ are the number density and thermal velocity of the 
buffer gas molecules, respectively. For our experiments, 
$\nu _n = 86\;{\mbox{s}}^{ - 1} \;$
 at the neon pressure of $20\;{\mbox{Pa}}$
 and $65\;{\mbox{s}}^{ - 1} \;$, at $15\;{\mbox{Pa}}$. 
The comparison with the above-discussed measurements of 
the damping coefficient $\nu$ shows that damping of the 
particle motion by the neutral drag must dominate but in 
this case, the wave extinction needs a more accurate 
treatment.

It is of interest to compare our results with those obtained 
from the Mach cone observations in 2D 
systems.\cite{26,21,30,27,28} The first difference is a 
single 3D Mach cone observed in this work (single cones 
were also observed in an argon discharge\cite{11,019,19}) 
vs.\ a double cone for a 2D system\cite{26,21} (in 
Refs.~\onlinecite{26} and \onlinecite{30}, three to four 
cones were observed). A double cone structure is 
sometimes associated with the propagation of the 
compressional and shear waves.\cite{28} While there is a 
principal possibility to excite the compressional and shear 
waves in a 2D system, the shear waves are unlikely to be 
observed in the 3D complex plasma. Indeed, in the vicinity 
of a projectile, a local melting of the dust crystal 
occurs,\cite{34} and the medium for the wave propagation 
becomes liquid. This rules out the shear waves and 
accounts for a single cone structure in our case. Note that 
there is a controversy about the number of the Mach cones 
in the 2D case and their nature as yet.

The second difference between the systems is the 
dissipation. In the experiments with 2D systems, the buffer 
gas pressure was an order of magnitude lower, and the 
particle radius, by several times larger than for our system 
(the particle diameter varied from about $5$
 to $9\;\mu {\mbox{m}}$
 in Ref.~\onlinecite{21}). Thus, in our case, the damping 
rate exceeds this quantity for the conditions of 
Ref.~\onlinecite{21} by almost two orders of magnitude. 
Apparently, a high damping is responsible for the absence 
of the interference patterns in our 3D case while these 
patterns are observed for a 2D system.\cite{21,28}

As for the effect of dissipation, the Epstein drag 
qualitatively accounts for the observed dependence $l(\nu 
)$
 in our case: the damping length decreases with the increase 
in $\nu$ (see the discussion above). Instead, in the 2D case, 
the opposite trend was registered.\cite{21}

A significant similarity between the two systems lies in the 
fact that solely the longest wavelength perturbation can be 
excited because $l \sim \lambda $, and both quantities have 
the scale of several interparticle distances. This allows one 
to conclude that for both systems, the observed Mach cones 
correspond to the nondispersive waves and define the speed 
of sound. It is noteworthy that the speed of sound 
determined in Ref.~\onlinecite{21} amounts to ca.\ 
$2\;{\mbox{cm/s}}$
 and is independent both on the sort of a gas (argon, xenon, 
krypton) and on the particle diameter. Almost the same 
speed of sound is listed in Table~\ref{t1}; likewise, $c_s$ 
is weakly dependent on the particle diameter. It is then 
surprising that for neon that was not investigated in the 
previous studies we obtained the speed of sound, which is 
twice as low as for other gases.

\section{\label{s5} Conclusion}

To summarize, we used the excitation of the Mach cones by 
large particles moving with the supersonic velocity for 
measurement of the speed of dust sound in a complex 
plasma with neon as a buffer gas. For this purpose, a 
high-definition method of the Mach cone visualization was 
developed. The determined speed of sound proved to be 
more than one order of magnitude lower than that predicted 
by the theory of the ion acoustic waves. We propose an 
interpretation of these results based on the similarity 
between a strongly coupled Coulomb system and a solid. 
Using a scaling law that relates the dust particle number 
density to its radius, we obtained the theoretical estimate 
for the speed of sound, which describes the main 
regularities of sound propagation in complex plasmas.

\begin{acknowledgments}

The authors gratefully acknowledge the support from the 
Russian Science Foundation (Project No.~14-12-01235) for 
theoretical interpretation of the experimental results and the 
support from DLR/BMWi (Grants Nos.~50WM0203 and 
50WM1203) for realizing our joint space experiments on 
the ISS.
\end{acknowledgments}
\providecommand{\noopsort}[1]{}\providecommand{\singleletter}[1]{#1}%
\end{document}